\documentstyle[preprint,aps,prb]{revtex}

\begin{document}

\draft
\preprint{Submitted to Phys.\ Rev.\ B}

\title{Singularity-matching peaks 
in superconducting single-electron transistor }

\author{Y. Nakamura$^1$, A. N. Korotkov$^{1,2}$, 
C. D. Chen$^1$, and J. S. Tsai$^1$}
\address{$^1$NEC Fundamental Research Laboratories, Tsukuba,
Ibaraki 305, Japan, \\
$^2$Nuclear Physics Institute, Moscow State University, 
Moscow 119899 GSP, Russia  }

\date{\today}

\maketitle

\begin{abstract}
We report the experimental observation of the recently predicted
peaks on the {\it I-V} curve of the superconducting single-electron
transistor at relatively high temperatures. The peaks are due to 
the matching of singularities in the quasiparticle density of 
states in two electrodes of a tunnel junction. The energy shift 
due to Coulomb blockade provides the matching at finite voltage.   

\end{abstract}


\vspace{1ex}

        Single-electron effects \cite{Av-Likh} 
in superconducting structures have several additional features 
\cite{Av-Likh,Tinkham}
in comparison with that in normal metals or semiconductors.
        The main differences are due to the specific role of the parity
of the electron number on a small island, \cite{Av-parity,Tuominen}
the effects of the Josephson coupling, 
\cite{Av-Likh,Tinkham,Fulton-JQP,Av-Al,Geerligs,Joyez}  
and the specific shape of the 
quasiparticle density of states (QDS). The last topic received 
relatively small attention
so far in both theoretical and experimental single-electronics,
although QDS leads to various interesting effects.
Besides the well-known shift of the Coulomb blockade threshold 
by $4 \Delta/e$ in SSS single-electron transistor (SET) 
(by $2\Delta /e$ in  NSN or 
SNS cases), let us mention the direct reproduction
of the QDS on the {\it I-V} curve of the SET with discrete energy spectrum
of the central electrode \cite{discr} and in the case of odd-parity 
current, \cite{Nakamura}
singularity-matching (SM) peaks at finite temperatures, \cite{Kor-SM}
the limitation of the differential conductance by quantum resistance
\cite{Kor-SM,Av-new} and the conductance jump at $V=4\Delta /e$ due 
to cotunneling \cite{Av-new}.

        In this paper we discuss the theory of SM peaks in more
detail and report their experimental observation
in the SSS SET. Somewhat similar experimental
results will be presented soon by another group. \cite{Pekola-SM}
 
        The origin of SM peaks can be easily understood from
the ``orthodox'' theory of the SET. \cite{Av-Likh,Likh-87}
Let us neglect the effects due to Josephson coupling and
consider only the quasiparticle tunneling. 
The dc current $I$ through the SET consisting of two tunnel junctions 
in series can be calculated from the equations \cite{Av-Likh,Likh-87}
        \begin{equation}
I=\sum_n  \left[ \Gamma_1^+ (n)-\Gamma_1^- (n) \right]  \sigma (n) \, , 
	\end{equation}
        \begin{equation}
    \sigma (n)  \left[ \Gamma_1^+ (n)+\Gamma_2^- (n)\right]  = 
\sigma (n+1)  \left[ \Gamma_1^- (n+1)+\Gamma_2^+ (n+1)\right]  ,
        \label{master-eq}\end{equation}
where $\Gamma_i^\pm (n)$ are the rates of tunneling through $i$th
junction ($i=1,2$) in the positive ($+$) or negative ($-$) direction when 
$n$ extra electrons are present on the central electrode of the SET, 
and $\sigma (n)$ is the probability of the charge state $n$.
In the case of $SS$ junction the tunneling rates
$\Gamma_i^\pm (n)\equiv \Gamma_i(W_i^\pm (n))$ are given by 
equations \cite{Tinkham}
                \begin{equation}
        \Gamma_i (W)=\frac{1}{e^2R_i} \int_{-\infty}^{+\infty}
                \rho (\varepsilon) f(\varepsilon ) 
        \rho (\varepsilon +W) [1-f(\varepsilon+W)] d\varepsilon ,
         \label{Gamma}\end{equation}
        \begin{equation}
        \rho(\varepsilon)=\sqrt{\frac{\varepsilon^2}{\varepsilon^2
        -\Delta(T)^2}} \, \theta (\varepsilon^2 -\Delta ^2) , 
         \end{equation}
where $\rho (\varepsilon )$ is the normalized QDS, $\Delta (T)$ is 
the superconducting energy gap, $\theta(x)$ is the step function, 
$f(\varepsilon)=1/(1+\exp (\varepsilon /T))$ is Fermi function, 
$T$ is temperature, $R_i$ is the normal tunnel resistance of $i$th 
junction, 
and the energy gain $W$ is given by \cite{Likh-87}
                \begin{equation}
        W_i^\pm (n)=\frac{e^2}{C_\Sigma} \left[ \pm\frac{VC_1C_2}{eC_i}
        -\frac{1}{2} \pm (-1)^i \left( n+\frac{Q_0}{e} \right) \right] .
        \label{W}\end{equation}
Here $C_1$ and $C_2$ are the junction capacitances, $C_\Sigma = C_1+C_2$  
is the total island capacitance, and $Q_0$ is the total 
background charge which accounts for the 
initial background charge $Q_{00}$ and the charge induced by 
the gate voltage $V_g$, $Q_0=Q_{00}+V_g C_g$ (for 
definiteness we consider the gate capacitance $C_g$ as being added to
$C_1$, although an arbitrary distribution of $C_g$ between $C_1$ and 
$C_2$ is possible in calculations \cite {Kor-Cg}).

        At low temperatures the quasiparticle current 
in SSS SET appears only above the voltage threshold 
        \begin{equation}
        V_t=\min_{i,n} (V_{i,n}^{QP} \,|\, V_t>4\Delta(T) )\, , 
        \label{V_t}\end{equation}
        \begin{equation}
V_{i,n}^{QP}=\frac{eC_i}{C_1C_2}\left[ \frac{2\Delta(T) C_\Sigma}{e^2}
+ \frac{1}{2}-(-1)^i \left( n+\frac{Q_0}{e} \right)\right] \, .
        \label{steps}\end{equation}
(The last equation is the condition $W^+_i(n)=2\Delta(T)$.)

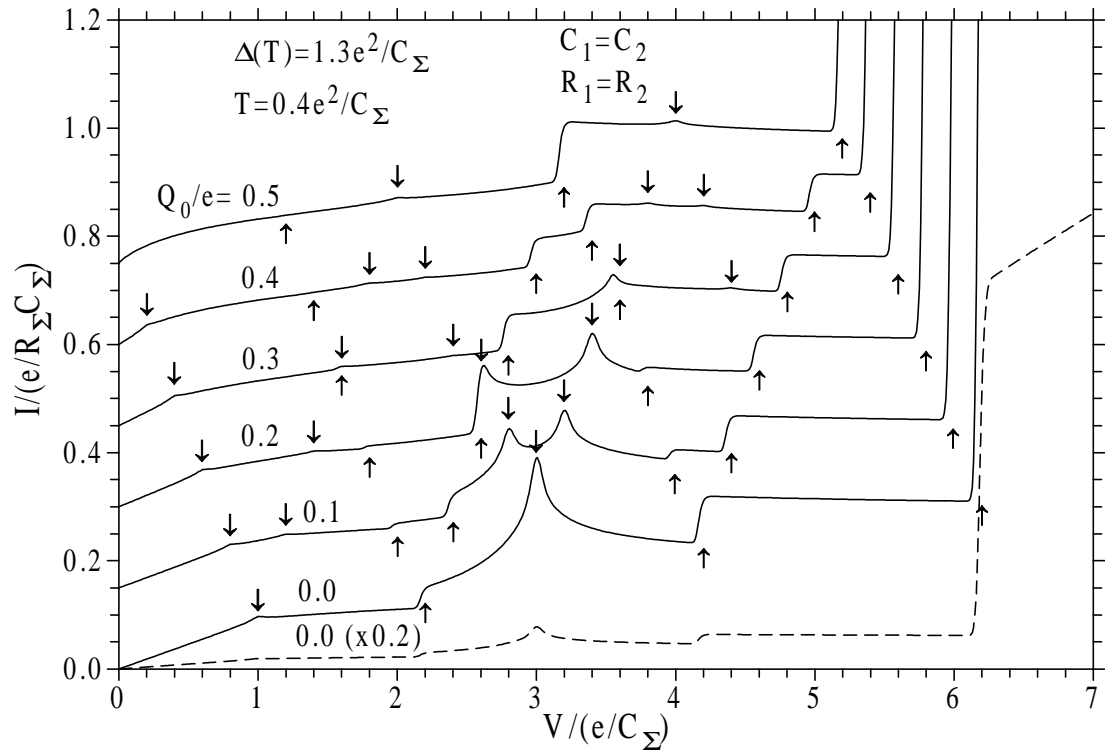
        \begin{figure}
\caption{ Theoretical {\it I-V} curves of the symmetrical SSS SET with 
$\Delta (T)=1.3 e^2/C_\Sigma$, $T=0.4 e^2/C_\Sigma$
at $V<V_t$ for several $Q_0$ taking into account only quasiparticle
tunneling. Notice the presence  
of SM peaks (marked by arrows down) and steps (arrows up).
Small phenomenological smearing $w=0.01 \Delta (T)$ of the superconducting
gap is assumed. Curves are shifted vertically for clarity.}
\label{peaks-steps}\end{figure}

        At temperatures $T$ comparable to the critical temperature $T_c$, 
the concentration of the thermally excited quasiparticles becomes
considerable, and this modifies the shape of the {\it I-V} curve,
in particular, creating several additional features 
at $V<V_t$.
   As an example, Fig.\ \ref{peaks-steps} shows the theoretical 
{\it I-V} curve of the symmetrical
SET-transistor with $\Delta (T)=1.3 e^2/C_\Sigma$, $T=0.4 e^2/C_\Sigma$ 
for several values of $Q_0$ (relatively large ratio $\Delta (T)/
(e^2/C_\Sigma)$ is chosen to show more features). 
One can see two types of features: peaks (marked by arrows down) 
and steps (arrows up).

        The peaks  positions constitute two series \cite{Kor-SM}
        \begin{equation}
        V_{i,n}^{SM}=\frac{eC_i}{C_1 C_2} \left( \frac{1}{2} -
        (-1)^i (n+\frac{Q_0}{e}) \right) ,
        \label{peaks}\end{equation}
        which correspond to the condition $W^+_i(n)=0$ (obviously, the
condition $W^-_i(n)=0$ gives the same set of voltages). For such 
tunneling with zero energy
gain the singularities of the density of states of two electrodes
match (remind that we consider the same $\Delta(T)$ in all electrodes),
that leads to the increase of the tunneling rate  $\Gamma^+_i(n)$
and explains the name of SM peaks.
In BCS
theory $\Gamma^+_i(n)$ is formally infinite at $W^+_i(n)=0$ (logarithmic
divergence).
        Although the current through SET-transistor remains finite
being governed by the stationary master equation  (\ref{master-eq}),
the divergence of $\Gamma$ would lead to very high and narrow center
of the SM peak. To take into account the inevitable smoothing of 
the singularity of $\rho (\varepsilon )$, in Fig.\ \ref{peaks-steps}
we assumed (phenomenologically) a small Gaussian inhomogeneous
broadening of $\Delta (T)$ with dispersion $w=0.01 \Delta (T)$.
The peak height depends very weakly (logarithmically) on $w$
provided $w \ll \Delta (T)$.
 
        The origin of SM peaks is similar to that of well-known peaks 
\cite{Tinkham} on
the {\it I-V} curve of a single junction with different energy gaps
$\Delta_1(T)$ and $\Delta_2(T)$ of electrodes at 
$V=[\Delta_1(T)-\Delta_2(T)]/e$. In our case the energy gaps can be 
the same, and the energy shift is provided by the Coulomb blockade.
        However, this analogy is not complete. For example, in our case
both singularities match simultaneously. Another
difference is that the reverse process (tunneling back) also has a 
large rate, and the net transport is due to the tunneling through
the other junction. 

        The voltage position of SM peaks coincides with the position
of the recently observed peaks \cite{Nakamura} in SSS SET at low 
temperatures when the parity-dependent current is due to the 
single quasiparticle created by the preceding tunneling event.  

        At not too high temperatures the SM peaks are more pronounced 
within the voltage interval 
$ 2\Delta(T)/e <V< 2\Delta(T)/e +e/C_\Sigma$ (see Fig.\ 
\ref{peaks-steps}). 
The lower bound is the
condition that the tunneling through the other junction which restores the
system into the initial charge state and gives the contribution to the
net current, is sufficiently fast, $W=eV>2\Delta(T)$.
The upper bound is the condition that after this restoring the tunneling
of the next electron through the same junction (which drives the system
out of ``resonance'') has a small rate, $W=eV-e^2/C_\Sigma<2\Delta(T)$.
        Hence, not more than two closely located peaks from the series
given by Eq.\ (\ref{peaks}) can be well pronounced on the {\it I-V} curve. 

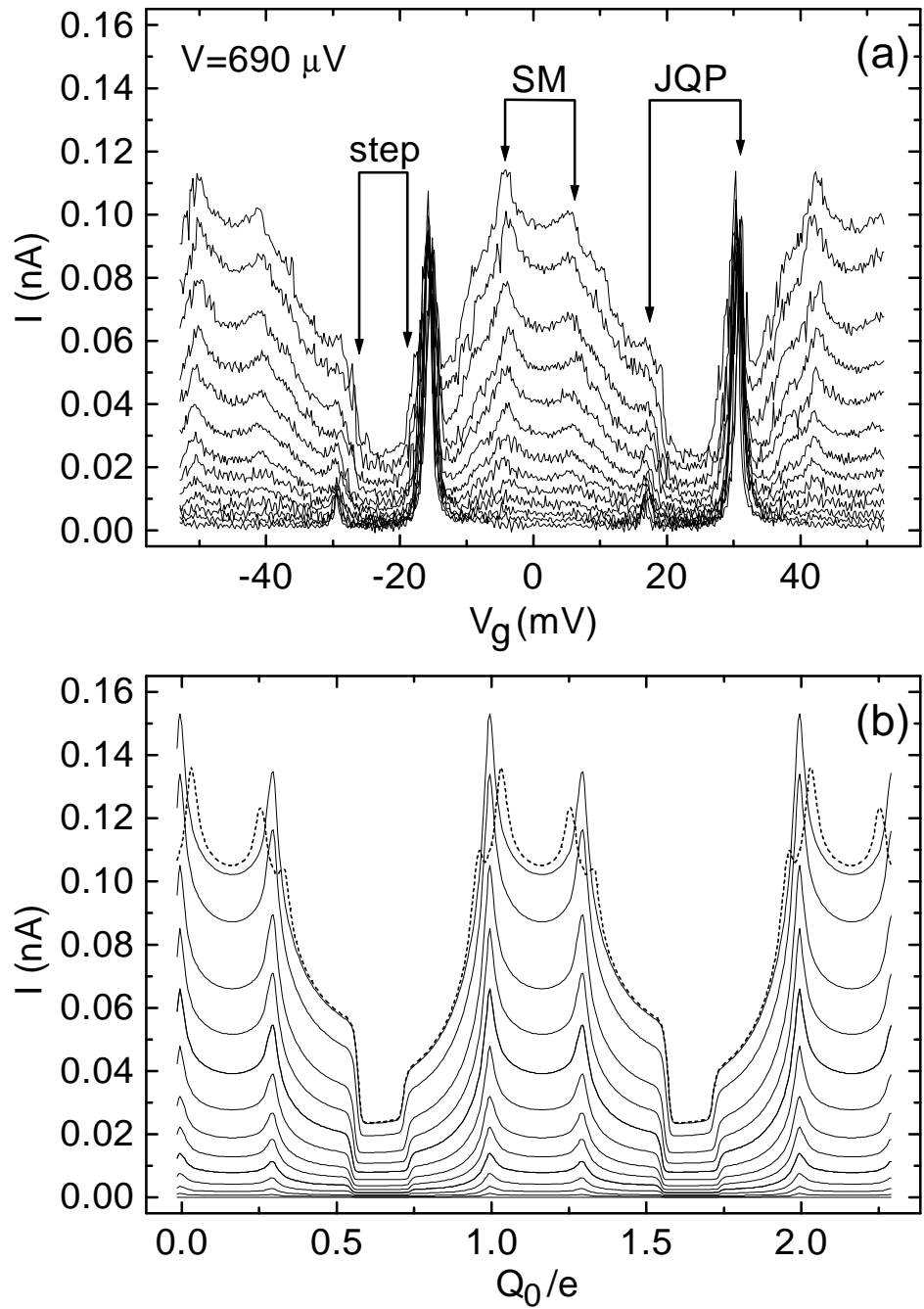
        \begin{figure}
\caption{(a) -- Experimental $I-V_g$ curves for $V=0.69$ mV at different 
temperatures which show two SM peaks (and also two JQP peaks and two
steps) per period. The temperatures in mK (from top to bottom) are: 
712, 684, 640, 605, 571, 532, 495, 462, 426, 386, 345, 303, 97. 
Notice that the height of SM peaks and steps grows with temperature.
(b) -- Solid lines show the corresponding theoretical prediction 
without fitting parameters.
The JQP peaks have not been included in simulations. 
Small smearing $w=0.03 \Delta (0)$ is assumed.
 Dashed line illustrates the peak splitting 
due to different $\Delta (T)$ in the island and leads ($\pm 10$ meV 
difference is used). }
\label{T-dep}\end{figure}

        The important property of SM peaks is their specific temperature 
dependence. They should be absent at small $T$ (because there are
no thermally excited quasiparticles), and their height grows
with $T$ for some temperature range (see Fig.\ \ref{T-dep}b below) 
until they begin to decrease
due to the suppression of superconductivity and/or correlation between
tunneling events. Notice that the voltage position of SM peaks does
not change with temperature despite the dependence $\Delta (T)$.

        One can see from Fig.\ \ref{peaks-steps} that the SM peaks
are rather broad and have asymmetric shape so that they have longer
and higher tail at the higher-voltage side. When the peak
is not well-pronounced, this tail resembles plateau.  
When the SM feature is even
weaker, it is seen as a small kink on the I-V curve (Fig.\ 
\ref{peaks-steps}).

        The other features seen in Fig.\ \ref{peaks-steps} 
are the step structures in the {\it I-V}
curve which are similar to the step at $V=V_t$. Their
positions satisfy the same condition $W^+_i(n)=2\Delta(T)$ 
and hence, the same Eq.\ (\ref{steps}) as for $V_t$.
        So the position of these two series 
of steps on $V-Q_0$ plane is just a continuation of the straight lines
corresponding to $V_t$ (they exist both above and below
$V_t$). 
        The steps in Fig.\ 
\ref{peaks-steps} are smoothed because of a finite $w$. 

        Notice that while the steps corresponding to Eq.\
(\ref{steps}) are usually positive (increase of the current), they
can also be negative -- for example, when the step position is
on the negative slope of an SM peak (the 
 decrease of the current occurs because 
the charge state having the resonant tunneling rate becomes less
probable). 

        Besides the steps described by Eq.\ (\ref{steps}), at relatively
high temperatures  the theory predicts an 
appearance of very small 
negative steps (both below and above $V_t$)  at
        \begin{equation}
V_{i,n}^{NS} = \frac{eC_i}{C_1C_2}\left[ -\frac{2\Delta(T) 
C_\Sigma}{e^2}
+\frac{1}{2}-(-1)^i(n+\frac{Q_0}{e})\right] ,
        \label{negsteps}\end{equation}
that corresponds to the condition  $W^+_i(n)=-2\Delta(T)$. The steps are 
negative because the dependence $\Gamma(W)$ given by Eq.\ (\ref{Gamma}) has
the step down at $W=-2\Delta(T)$. The effect is very weak because of the 
factor
$\exp (-4\Delta(T)/T)$ and also because of the existence of the
opposite effect due to the simultaneous threshold condition
$W^-_i(n-(-1)^i)=2\Delta(T)$  
(negative steps have not been observed in our experiment).

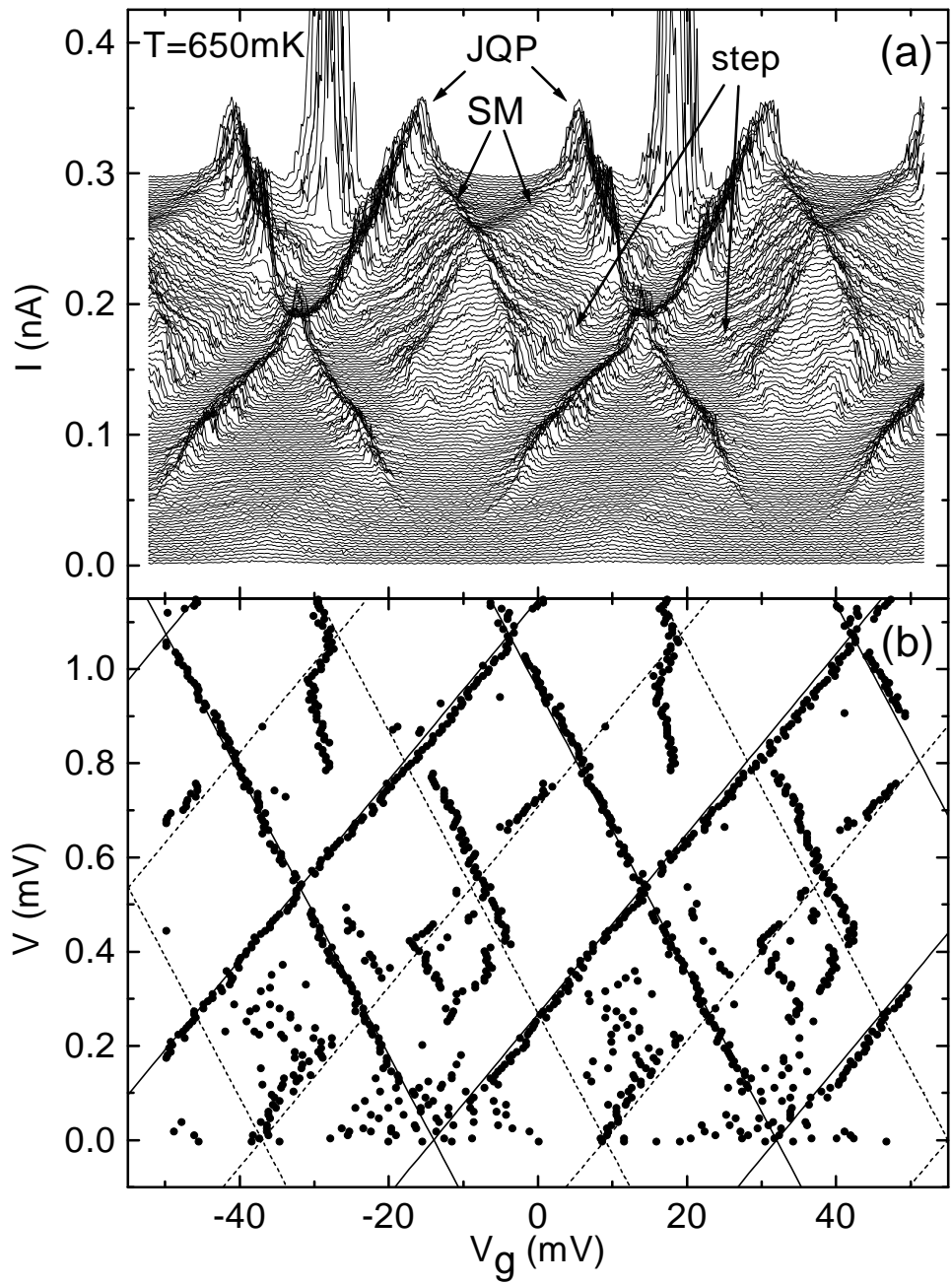
        \begin{figure}
\caption{ (a) -- The experimental dependence of the current $I$ 
on the gate 
voltage $V_g$ for SSS SET at $T=650$ mK. The bias voltages $V$ range 
from 0 to 0.828 mV with the step of 7.08 $\mu$V. The curves are 
shifted vertically by 
$\Delta I $(pA)$=281\times V$(mV). 
(b) -- The positions of the current maxima 
 on $V-V_g$ plane. The solid lines fit JQP peaks. 
The dashed lines show theoretical position of SM peaks.}
\label{maxima}\end{figure}

        The aluminum-based single-electron transistors were fabricated 
using the standard
two-angle evaporation technique. The details of the fabrication are
given in Ref. \cite{JQP-pos}.
        Figure \ref{maxima}a shows the experimental dependence of 
the current on the gate voltage for different bias voltages
at $T=650$ mK. 
The largest feature seen is the Figure is the onset of the
fast quasiparticle tunneling at $V>V_t$ (Eq.\ (\ref{V_t})).   
(Actually, we see peaks because of the small current scale of 
the Fig.\ \ref{maxima}a; for larger bias voltages they 
transform into plateaus with sufficiently sharp edges.) 
The well-pronounced peaks along the straight lines intersecting
the abscissa axis at $V_g=-14$mV and $V_g=32$mV are JQP 
peaks.\cite{Fulton-JQP} (They are due to Josephson coupling and, hence,
are outside of the scope of the present paper. 
The position of JQP peaks is 
given\cite{Fulton-JQP,Av-Al,JQP-pos} by  Eq.\ (\ref{peaks}) 
without the term 1/2 inside parentheses.)
The step structures can be seen along the lines (see Eq.\ 
(\ref{steps})) which are 
the continuation of the main threshold $V_t(V_g)$ (they start 
from the large features due to $V_t$ in the upper part of Fig.\  
\ref{maxima}a).
 And finally, the SM peaks are represented as 
rather broad features along the strait lines approximately 
in the middle (theoretically exactly in the middle) 
between JQP lines which intersect the abscissa axis roughly
at $V_g\simeq 10$mV and $V_g\simeq -40$mV.
 Small SM peaks have been observed even above $V_t$. 
 The SM features along the
line with negative slope are more pronounced than along
the line with positive slope possibly because of the difference
in junction resistances. 
 Similar measurements made at $T=50$ mK
do not show SM features as well as additional step structures while
JQP peaks remain well-pronounced at $V \agt 0.65$ mV.

        Figure \ref{maxima}b shows the numerically determined  
positions of the maxima of the $I-V_g$ 
curves from Fig.\ \ref{maxima}a on $V-V_g$ plane. 
From the straight lines corresponding to JQP peaks (solid lines) 
we determine
the junction capacitances $C_1=183\pm 4$ aF, $C_2=117\pm 3$ aF.  
The gate capacitance $C_g=3.5$ aF which determines the gate voltage 
period of 46 mV is included into $C_1$ because $V_g$ has been measured
from the outer side of $C_1$. Notice that the bias voltage 
corresponding to the intersection of two JQP lines directly gives 
the charging energy, $e/C_\Sigma =0.53$ mV. The minimal $V_t$ of 
0.80 mV is used to determine the superconducting gap 
$\Delta (T)=0.20$ meV 
(minimal $V_t$ corresponds to the edge of
the almost vertical curved lines in the upper part of Fig.\ 
\ref{maxima}b). 

        Dashed lines in Fig.\ \ref{maxima}b show the theoretical
position $V_{i,n}^{SM}$ of SM peaks calculated from 
Eq.\ (\ref{peaks}). We see
that experimental peaks are located at somewhat higher bias voltages.
This can be explained by several reasons. 
	First, SM feature has a rather smooth shape, and, hence,  
the addition of any current component which 
increases with bias voltage leads to the apparent shift of the 
maximum to higher voltages (we also checked numerically that 
relatively large inhomogeneous broadening of $\Delta (T)$ leads 
to a similar shift). 
	Second, the additional 
contribution to the position shift in Fig.\ \ref{maxima}b can occur 
because the peaks are determined as the maximum current point 
over $V_g$, not over $V$. 
 (The $V_g$ change which decreases $V_{i,n}^{SM}$ 
also weakens Coulomb blockade in the same junction, hence,
increasing the ``background'' current and leading to the apparent
shift of the maximum position.) Finally, the third possible 
explanation of the shift (which we believe is most likely) is due 
to the difference between $\Delta (T)$ in the island and leads. 
Then each SM peak should split in two (there is some experimental
evidence of such a splitting which is slightly seen in Fig.\ 
\ref{maxima}a and Fig.\ \ref{T-dep}a). 
Numerical simulations show that the peak corresponding to higher 
bias voltage is more pronounced (see Fig.\ \ref{T-dep}b) while 
the lower peak is possibly too small to be represented in Fig.\ 
\ref{maxima}b. 
The difference about $0.02$ meV between the energy gaps would be 
sufficient to explain the experimental deviation.

        The experimental temperature dependence of SM peaks
on $I-V_g$ curves
is shown in Fig.\ \ref{T-dep}a. Besides two SM peaks per period 
of $V_g$ one can see two steps and two JQP peaks (the heights of
JQP peaks are different because the Josephson coupling in one
junction has been suppressed\cite{fabricat}). We see that SM peaks 
as well as steps   grow with temperature. Solid lines in Fig.\ 
\ref{T-dep}b show the
corresponding theoretical $I-Q_0$ curves calculated without 
fitting parameters (JQP peaks are not included in the model). 
The total resistance $R_\Sigma = 605$ k$\Omega$ 
is obtained from the $I-V$ curve and $R_1/R_2=C_2/C_1$ is 
assumed.
 The gap $\Delta (0)=0.207$ meV is used to get  
$\Delta (T)= 0.2$ mV at $T=650$ mK.
        The small broadening $w=0.03\Delta (0)$ of the gap was used
to eliminate the unphysical divergence of $\Gamma$ at the peak center,
while we did not attempt to fit experimental SM peak height by $w$
(larger $w$ decreases the height, though the dependence is quite weak 
for $w \alt 0.05 \Delta (T)$).
 The good correspondence
between Figs.\ \ref{T-dep}a and \ref{T-dep}b is an additional
proof that the observed peaks are really SM peaks.
	The dashed curve (corresponding to the top solid curve) 
illustrates the peak splitting due to different $\Delta (T)$ 
in the island and leads (difference of 20 meV is used). One can see that 
this assumption not only explains the peak position shift 
and traces of such a splitting in experiment, but also improves
the agreement for the peak height. 

        Let us mention that the height of thermally activated 
JQP peak (at $V<2\Delta /e+e/2C_\Sigma$) as a function of $V$ 
($V_g$ is varied correspondingly) should also exhibit 
the SM feature at 
$V=e/2C_\Sigma$ because at this voltage the tunneling of the 
second quasiparticle in JQP cycle is at resonance. 
	There are some traces of such an increase in Fig.\
\ref{maxima}a and also  one can see qualitatively 
that in Fig.\ \ref{maxima}a the JQP peaks start to
decrease crudely at $V<e/2C_\Sigma$ (because of the thermal 
broadening of SM feature this smooth boundary moves to
lower voltages by $\delta V \simeq TC_\Sigma /C_i \simeq 0.1 $ mV).  
        Similar behavior should be expected for the height of
thermally activated steps with SM feature at $V=2 \Delta (T)$. 
         
        In conclusion, we observed SM peaks on $I-V_g$ dependence
of SSS SET at temperatures comparable to $T_c$. The shape and
position of the features agree well with the theoretical prediction.

        We thank the group from the University of Jyv\"{a}skyl\"{a} 
for the information about their experimental results 
\cite{Pekola-SM} prior to publication. 
The work was partially performed under the management of FED
as a part of the MITI R\&D Program Superconducting Electron Devices 
Project supported by NEDO.

\end{document}